\documentclass[11pt]{article}
\begin{document}

\begin{center}
\large\textbf{Dirac particles' tunnelling from 5-dimensional rotating black strings influenced by the generalized uncertainty principle}
\end{center}

\begin{center}
Deyou Chen \footnote{E-mail: \underline{dyouchen@gmail.com} }
\end{center}

\begin{center}
Institute of Theoretical Physics, China West Normal University, \\Nanchong 637009, China
\end{center}

\textbf{Abstract:} The standard Hawking formula predicts the complete evaporation of black holes. Taking into account effects of quantum gravity, we investigate fermions' tunnelling from a 5-dimensional rotating black string. The temperature is determined not only by the string, but also affected by the quantum number of the emitted fermion and the effect of the extra spatial dimension. The quantum correction slows down the increase of the temperature, which naturally leads to the remnant in the evaporation.

\section{Introduction}

The semi-classical tunnelling method is an effective way to describe the Hawking radiation \cite{KW}. Using this method, the tunnelling behavior of massless particles across the horizon was veritably described in \cite{PW}. In the research, the varied background spacetime was taken into account. The tunnelling rate was related to the change of the Bekenstein-Hawking entropy and the temperature was higher than the standard Hawking temperature. In the former researches, the standard temperatures were derived \cite{SWH,DR,WGU,RW,IUW,MP}, which imply the complete evaporation of black holes. Thus the varied background spacetime accelerates the black holes' evaporation. This result was also demonstrated in other complicated spacetimes \cite{ECV,AMV,WJ,SZY,HZZ}. Extended this work to massive particles, the tunnelling radiation of general spacetimes was investigated in \cite{ZZ,JWC}. The same result was derived by the relation between the phase velocity and the group velocity.

In \cite{KM}, the standard Hawking temperatures were recovered by fermions tunnelling across the horizons. In the derivation, the action of the emitted particle was derived by the Hamilton-Jacobi equation \cite{ANVZ}. This derivation is based on the complex path analysis \cite{KP}. In this method, we don't need the consideration of that the particle moves along the radial direction \cite{LR,CV,QQJ,LY}. This is a difference from the work of Parikh and Wilczek \cite{PW}.

The tunnelling radiation beyond the semi-classical approximation was discussed in \cite{BM,BRM,SVZR}. Their ansatz is also based on the Hamilton-Jacobi method. The key point is to expand the action in a powers of $\hbar$. Using the expansion, one can get the quantum corrections over the semiclassical value. The corrected temperature is lower than the standard Hawking temperature. The higher order correction entropies were derived by the first law of black hole
thermodynamics.

Taking into account effects of quantum gravity, the semi-classical tunnelling method was reviewed in the recent work \cite{NS,CWY}. In \cite{NS}, the tunnelling of massless particles through quantum horizon of a Schwarzschild black hole was investigated by the influence of the generalized uncertainty principle (GUP). Through the modified commutation relation between the radial coordinate and the conjugate momentum and the deformed Hamiltonian equation, the radiation spectrum with the quantum correction was derived. The thermodynamic quantities were discussed. In the fermionic fields, with the consideration of effects of quantum gravity, the generalized Dirac equation in curved spacetime was derived by the modified fundamental commutation relations \cite{KMM}, which is \cite{CWY}

\begin{eqnarray}
&&\left[i\gamma^{0}\partial_{0}+i\gamma^{i}\partial_{i}\left(1-\beta m^{2}
\right)+i\gamma^{i}\beta\hbar^{2}\left(\partial_{j}\partial^{j}\right)
\partial_{i}+\frac{m}{\hbar}\left(1+\beta\hbar^{2}\partial_{j}\partial^{j}
-\beta m^{2}\right)\right.\nonumber \\
&&\left.+i\gamma^{\mu}\Omega_{\mu}\left(1+\beta\hbar^{2}\partial_{j}\partial^{j}
-\beta m^{2}\right)\right]\psi = 0.
\label{eq1.1}
\end{eqnarray}

\noindent This derivation is based on the existence of a minimum measurable length. The length can be realized in a model of GUP

\begin{eqnarray}
\Delta x \Delta p \geq \frac{\hbar}{2}\left[1+ \beta (\Delta p)^2+ \beta <p>^2\right],
\label{eq1.2}
\end{eqnarray}

\noindent where $\beta = \beta_0 \frac{l^2_p}{\hbar^2}$ is a small value, $\beta_0 <10^{34}$ is a dimensionless parameter and $l_p$ is the Planck length. Eq. (\ref{eq1.2}) was derived by the modified Heisenberg algebra $\left[x_i,p_j\right]= i \hbar \delta_{ij}\left[1+ \beta p^2\right]$, where $x_i$ and $p_i$ are position and momentum operators defined respectively as  \cite{KMM,DV}

\begin{eqnarray}
x_i &=& x_{0i}, \nonumber\\
p_i &=& p_{0i} (1 + \beta p_0^2),
\label{eq1.3}
\end{eqnarray}

\noindent $p_0^2 = \sum p_{0j}p_{0j}$, $x_{0i}$ and $p_{0j}$ satisfy the canonical commutation relations $\left[x_{0i},p_{0j}\right]= i \hbar \delta_{ij}$. Thus the minimal position uncertainty is gotten as

\begin{eqnarray}
\Delta x &=& \hbar \sqrt{\beta} \sqrt{1+ \beta <p>^2} ,
\label{eq1.4}
\end{eqnarray}

\noindent which means that the minimum measurable length is $ \Delta x_0 = \hbar \sqrt{\beta}$ \cite{KMM}. To let $\Delta x_0 $ have a physical meaning, the condition $\beta >0$ must be satisfied. It was showed in \cite{KMM}. Based on the GUP, the black hole's remnant was first researched by Adler et al. \cite{ACS}. Incorporate eq. (\ref{eq1.3}) into the Dirac equation in curved spacetime, the modified Dirac equation was derived\cite{CWY}. Using this modified equation, fermions' tunnelling from the Schwarzschild spacetime was investigated. The temperature was showed to be related to the quantum number of the emitted fermion. An interested result is that the quantum correction slows down the increase of the temperature. It is natural to lead to the remnant.

In this paper, taking into account effects of quantum gravity, we investigate fermions' tunnelling from a 5-dimensional rotating black string. The key point in this paper is to construct a tetrad and five gamma matrices. The result shows that in the frame of quantum gravity, the temperature is affected not only by the quantum number of the emitted fermion, but also by the effect of the extra compact dimension. The quantum correction slows down the increase of the temperature. The remnant is naturally observed in the evaporation.

In the next section, we perform the dragging coordinate transformation on the metric and construct five gamma matrices, then investigate the fermion's tunnelling from the 5-dimensional rotating string. The remnant is observed. Section 3 is devoted to our conclusion.

\section{Tunnelling radiation with the influence of the generalized uncertainty principle}

The Kerr metric describes a rotating black hole solution of the Einstein equations in four dimensions. When we add an extra compact spatial dimension to it, the metric becomes

\begin{eqnarray}
ds^{2} &=& -\frac{\Delta}{\rho^2}\left(dt - a\sin^2\theta d\varphi\right)^2+ \frac{\sin^2\theta}{\rho^2}\left[adt - (r^2+a^2)d\varphi\right]^2 \nonumber\\
&& +\frac{\rho^2}{\Delta}dr^2 +\rho^2d\theta^2+ g_{zz}dz^2,
\label{eq2.1}
\end{eqnarray}

\noindent where $\Delta= r^2-2Mr +a^2 = (r-r_+)(r-r_-)$, $\rho^2  = r^2+a^2\cos^2\theta$, $g_{zz}$ is usually set to $1$. The above metric describes a rotating uniform black string. $r_{\pm}= M\pm \sqrt{M^2-a^2}$ are the locations of the event (inner) horizons. $M$ and $a$ are the mass and angular momentum unit mass of the string, respectively. A fermion's motion satisfies the generalized Dirac equation (\ref{eq1.1}). To investigate the tunnelling behavior of the fermion, it can directly choose a tetrad and construct gamma matrices from the metric (\ref{eq2.1}). The metric (\ref{eq2.1}) describes a rotating spacetime. The energy and mass near the horizons are dragged by the rotating spacetime. It is not convenient to discuss the fermion's tunnelling behavior. For the convenience of constructing the tetrad and gamma matrices, we perform the dragging coordinate transformation $d\phi = d\varphi - \Omega dt $, where

\begin{equation}
\Omega =\frac{\left(r^2+a^2- \Delta\right) a}{\left( r^2+a^2\right)^2 -\Delta a^2\sin^2\theta},
\label{eq2.2}
\end{equation}

\noindent on the metric (\ref{eq2.1}). Then the metric (\ref{eq2.1}) takes on the form as

\begin{eqnarray}
ds^2 &=& - F(r)dt^2+\frac{1}{G(r)}dr^{2} +g_{\theta \theta}d\theta^2 +g_{\phi\phi} d\phi^2 + g_{zz}dz^2\nonumber\\
&=& - \frac{\Delta \rho^2}{(r^2+a^2)^2-\Delta a^2 \sin^2{\theta}}dt^2 +\frac{\rho^{2}}{\Delta}dr^{2} + g_{zz}dz^2\nonumber\\
&&+\rho^2 d\theta^2 + \frac{\sin^2{\theta}}{\rho^2}\left[(r^2+a^2)^2-\Delta a^2 \sin^2{\theta}\right]d\phi^2.
\label{eq2.3}
\end{eqnarray}

\noindent Now the tetrad is directly constructed from the above metric. It is

\begin{eqnarray}
e_{\mu}^a= diag(\sqrt{F},1/\sqrt{G},\sqrt{g_{\theta\theta}},\sqrt{g_{\phi\phi}},\sqrt{g_{zz}}).
\label{eq2.4}
\end{eqnarray}

\noindent Then gamma matrices are easily constructed as follows

\begin{eqnarray}
\gamma^{t}=\frac{1}{\sqrt{F}}\left(\begin{array}{cc}
0 & I\\
-I & 0
\end{array}\right), &  & \gamma^{\theta}=\sqrt{g^{\theta\theta}}\left(\begin{array}{cc}
0 & \sigma^{2}\\
\sigma^{2} & 0
\end{array}\right),\nonumber \\
\gamma^{r}=\sqrt{G}\left(\begin{array}{cc}
0 & \sigma^{3}\\
\sigma^{3} & 0
\end{array}\right), &  & \gamma^{\phi}=\sqrt{g^{\phi\phi}}\left(\begin{array}{cc}
0 & \sigma^{1}\\
\sigma^{1} & 0
\end{array}\right),\nonumber \\
\gamma^{z}=\sqrt{g^{zz}}\left(\begin{array}{cc}
-I & 0\\
0 & I
\end{array}\right).
\label{eq2.5}
\end{eqnarray}

\noindent When measure the quantum property of a spin-1/2 fermion, we can get two values. They correspond to two states with spin up and spin down. The wave functions of two states of a fermion in the metric (\ref{eq2.3}) spacetime take on the form as

\begin{eqnarray}
\psi_{\left(\uparrow\right)}=\left(\begin{array}{c}
A\\
0\\
B\\
0
\end{array}\right)\exp\left(\frac{i}{\hbar}I_{\uparrow}\left(t,r,\theta,\phi,z\right)\right),
\label{eq2.6}
\end{eqnarray}

\begin{eqnarray}
\psi_{\left(\downarrow\right)}=\left(\begin{array}{c}
0\\
C\\
0\\
D
\end{array}\right)\exp\left(\frac{i}{\hbar}I_{\downarrow}\left(t,r,\theta,\phi,z\right)\right),
\label{eq2.7}
\end{eqnarray}

\noindent where $A, B, C, D$ are functions of $(t, r, \theta, \phi, z)$, and $I$ is the action of the fermion, $\uparrow$ and $\downarrow$ denote the spin up and spin down, respectively. In this paper, we only investigate the state with spin up. The analysis of the state with spin down is parallel. To use the WKB approximation, we insert the wave function (\ref{eq2.6}) and the gamma matrices into the generalized Dirac equation (\ref{eq1.1}). Dividing by the exponential term and considering the leading terms yield four equations. They are

\begin{eqnarray}
-\frac{B}{\sqrt{F}}\partial_t I- B \sqrt{G}(1-\beta m^2)\partial_r I + A \sqrt{g^{zz}}(1-\beta m^2)\partial_z I \nonumber\\
- A m(1-\beta m^2- \beta Q)+ B\beta \sqrt{G} Q \partial_r I - A\beta \sqrt{g^{zz}} Q \partial_z I = 0,
\label{eq2.8}
\end{eqnarray}

\begin{eqnarray}
\frac{A}{\sqrt{F}}\partial_t I- A \sqrt{G}(1-\beta m^2)\partial_r I -B \sqrt{g^{zz}}(1-\beta m^2)\partial_z I \nonumber\\
- B m(1-\beta m^2- \beta Q)+ A\beta \sqrt{G} Q \partial_r I +B\beta \sqrt{g^{zz}} Q \partial_z I = 0,
\label{eq2.9}
\end{eqnarray}

\begin{eqnarray}
-B\left(i\sqrt{g^{\theta\theta}}\partial_{\theta} I + \sqrt{g^{\phi\phi}}\partial_{\phi} I \right)(1-\beta m^2 - \beta Q) = 0,
\label{eq2.10}
\end{eqnarray}

\begin{eqnarray}
-A\left(i\sqrt{g^{\theta\theta}}\partial_{\theta} I + \sqrt{g^{\phi\phi}}\partial_{\phi} I \right)(1-\beta m^2 - \beta Q) = 0,
\label{eq2.11}
\end{eqnarray}

\noindent where $Q= g^{rr} \left( {\partial _r I} \right)^2  + g^{\theta \theta } \left( {\partial _\theta  I} \right)^2 + g^{\phi \phi } \left( {\partial _\phi  I} \right)^2+ g^{zz} \left( {\partial_z I} \right)^2$. It is difficult to get the expression of the action from the above equations. Considering the property of the spacetime, we carry out separation of variables as

\begin{eqnarray}
I = -(\omega-j\Omega)t + W(r) + \Theta (\theta,\phi) +Jz,
\label{eq2.12}
\end{eqnarray}

\noindent where $\omega$ is the energy of the emitted fermion, $j$ is the angular momentum and $J$ is a conserved momentum corresponding to the compact dimension. Eqs. (\ref{eq2.10}) and (\ref{eq2.11}) are irrelevant to $A, B$. Inserting Eq. (\ref{eq2.12}) into them yields

\begin{eqnarray}
i\sqrt{g^{\theta\theta}}\partial_{\theta} \Theta + \sqrt{g^{\phi\phi}}\partial_{\phi} \Theta = 0,
\label{eq2.13}
\end{eqnarray}

\noindent which implies that $\Theta$ is a complex function other than the constant solution. In the former research, it was found that the contribution of $\Theta$ could be canceled in the derivation of the tunnelling rate. Using Eq. (\ref{eq2.13}), an important relation is easily gotten as

\begin{eqnarray}
g^{\theta\theta}(\partial_{\theta} \Theta)^2 + g^{\phi\phi}(\partial_{\phi} \Theta)^2 = 0.
\label{eq2.14}
\end{eqnarray}

\noindent Now our interest is the first two equations. Inserting Eq. (\ref{eq2.12}) into Eqs. (\ref{eq2.8}) and (\ref{eq2.9}), canceling $A$ and $B$ and neglecting the higher order terms of $\beta$, we get

\begin{eqnarray}
A(\partial_r W)^4 + B(\partial_r W)^2 + C = 0,
\label{eq2.15}
\end{eqnarray}

\noindent where

\begin{eqnarray}
A &=& 2\beta G^2F,\nonumber\\
B &=& -[1-4\beta g^{zz} \left( {\partial_z I} \right)^2]GF,\nonumber\\
C &=& [1-2\beta m^2-2\beta g^{zz}\left( {\partial_z I} \right)^2](m^2- g^{zz}\left( {\partial_z I} \right)^2)F+\left( {\partial_t I} \right)^2.
\label{eq2.16}
\end{eqnarray}

\noindent Solving the above equation at the event horizon yields the imaginary part of the radial action. Based on the invariance under canonical transformations, we adopt the method developed in \cite{APAS}. The tunnelling rate is

\begin{eqnarray}
\Gamma &\propto & exp[-\frac{1}{\hbar}Im\oint p_r dr] =exp\left[-\frac{1}{\hbar}Im\left(\int p_r^{out}dr-\int p_r^{in}dr\right)\right]\nonumber\\
&=& exp\left[\mp \frac{2}{\hbar}Im\int p_r^{out,in}dr\right] .
\label{eq2.17}
\end{eqnarray}

\noindent In the above equation, $\oint p_r dr$ is invariant under canonical transformations. Here let $p_r = \partial_rW$. Thus the solutions of $Im\int p_r^{out,in}dr$ are determined by Eq. (\ref{eq2.15}), which is

\begin{eqnarray}
Im\oint p_r dr &=&  2Im W^{out}\nonumber\\
&=& 2 Im\int dr\sqrt{\frac{(E-j\Omega)^2+(1-2\beta m^2 -2\beta g^{zz}  J^2)(m^2-g^{zz} J^2)F}{GF(1-4\beta g^{zz} J^2)}}\nonumber\\
&& \times \left[1+ \beta \left(\frac{(E-j\Omega)^2}{F}+m^2-g^{zz} J^2\right)\right]\nonumber\\
&=& 2 \pi \frac{(\omega -j\Omega_+)(r_+^2 +a^2)}{r_+ - r_-}\left[1+\beta \Xi(J,\theta,r_+,j)\right],
\label{eq2.18}
\end{eqnarray}

\noindent where $g^{zz} =1$, $\Omega_+ = \frac{a}{r_+^2 +a^2}$ is the angular velocity at the event horizon. $\Xi(J,\theta,r_+,j) $ is a complicated function of $J,\theta,r_+,j$, therefore, we don't write down here. It should be that  $\Xi(J,\theta,r_+,j) >0$. If adopt Eq. (\ref{eq2.18}) to calculate the tunnelling rate, we will derive two times Hawking temperature, which was showed in \cite{AAPS}. This is not in consistence with the standard temperature. With careful observations, Akhmedova et. al. found that the contribution coming from the temporal part of the action was ignored \cite{APAS}. When they took into account the temporal contribution, the factor of two in the temperature was resolved.

To find the temporal contribution, we use the Kruskal coordinates $(T,R)$. The region exterior to the string $(r>r_+)$ is described by

\begin{eqnarray}
T &=& e^{\kappa_+r_*}sinh(\kappa_+t),\nonumber\\
R &=& e^{\kappa_+r_*}cosh(\kappa_+t),
\label{eq2.19}
\end{eqnarray}

\noindent where $r_*= r+ \frac{1}{2\kappa_+}ln\frac{r-r_+}{r_+}- \frac{1}{2\kappa_-}ln\frac{r-r_-}{r_-}$ is the tortoise coordinate, and $\kappa_\pm =\frac{r_+-r_-}{2(r_{\pm}^2+a^2)}$ denote the surface gravity at the outer (inner) horizons. The description of the interior region is given by

\begin{eqnarray}
T &=& e^{\kappa_+r_*}cosh(\kappa_+t),\nonumber\\
R &=& e^{\kappa_+r_*}sinh(\kappa_+t).
\label{eq2.20}
\end{eqnarray}

\noindent To connect these two patches across the horizon, we need to rotate the time $t$ as $t\rightarrow t-i\kappa_+\frac{\pi}{2}$. As pointed in \cite{APAS}, this ¡°rotation¡± would lead to an additional imaginary contribution coming from the temporal part, namely, $Im(E\Delta t^{out,in})=\frac{1}{2}\pi E\kappa_+$, where $E=\omega - j \Omega_+$. Thus the total temporal contribution is $Im(E\Delta t)=\pi E\kappa_+$.  Therefore, the tunnelling rate is

\begin{eqnarray}
\Gamma &\propto & exp\left[-\frac{1}{\hbar}\left(Im (E\Delta t)+Im\oint p_r dr\right)\right] \nonumber\\
&=& -4\pi \frac{(\omega -j\Omega_+)(r_+^2 +a^2)}{\hbar(r_+ - r_-)}\left[1+\frac{1}{2}\beta \Xi(J,\theta,r_+,j)\right].
\label{eq2.21}
\end{eqnarray}

\noindent This is the Boltzman factor expression and implies the temperature

\begin{eqnarray}
T &=& \frac{\hbar(r_+ - r_-)}{4\pi (r_+^2 +a^2)\left[1+\frac{1}{2}\beta \Xi(J,\theta,r_+,j)\right]}\nonumber\\
&=& T_0\left[1-\frac{1}{2}\beta \Xi(J,\theta,r_+,j)\right],
\label{eq2.22}
\end{eqnarray}

\noindent where $T_0=  \frac{\hbar(r_+ - r_-)}{4\pi (r_+^2 +a^2)}$ is the standard Hawking temperature of the Kerr string and shares the same expression of the temperature of the 4-dimensional Kerr black hole. It shows that the corrected temperature is determined by the mass, angular momentum and extra dimension of the string, but also affected by the quantum number (energy, mass and angular momentum) of the fermion. Therefore, the property of the emitted fermion affects the temperature when the effects of quantum gravity are taken into account.

When $a=0$, the metric (\ref{eq2.1}) is reduced to the Schwarzschild string metric. Then the imaginary part of the radial action (\ref{eq2.18}) is reduced to

\begin{eqnarray}
Im\oint p_r dr&=& 2 \pi \omega r_+\left[1+\beta \left(2\omega^2 +3m^2/2+J^2/2\right)\right].
\label{eq2.23}
\end{eqnarray}

\noindent Adopting the same process, we get the temperature of the Schwarzschild string as

\begin{eqnarray}
T &=& \frac{\hbar}{4\pi r_+\left[1+\beta \left(\omega^2 +3m^2/4+J^2/4\right)\right]}\nonumber\\
&=& \frac{\hbar}{8\pi M}\left[1-\beta \left(\omega^2 +3m^2/4+J^2/4\right)\right].
\label{eq2.24}
\end{eqnarray}

\noindent It shows that the effect of the extra dimension and the quantum number (energy, mass and angular momentum) of the fermion affect the temperature of the Schwarzschild string. It is obviously that the quantum correction slows down the crease of the temperature. Finally, the string can not evaporate completely and there is a blanched state. At the this state, the remnant is left. The effect of the extra dimension plays an role of impediment during the evaporation. When $J = 0$, Eq. (\ref{eq2.24}) describes the temperature of the 4-dimensional Schwarzschild black hole. The remnant was derived as $\geq M_p/{\beta_0}$, where $M_p$ is the Planck mass and $\beta_0$ is a dimensionless parameter accounting for quantum gravity effects \cite{CWY}.

\section{Conclusion}

In this paper, we investigated the fermion's tunnelling from the 5-dimensional Kerr string spacetime. To incorporate the influence of quantum gravity, we adopted the generalized Dirac equation derived in \cite{CWY}. The corrected temperature is not only determined by the mass, angular momentum and extra dimension, but also affected by the quantum number of the emitted fermion. The quantum correction slows down the increase of the temperature. Finally, the balance state appears. At this state, the string can not evaporate completely and the remnant is left. This can be seen as the direct consequence of the generalized uncertainty principle.

\vspace*{3.0ex}
{\bf Acknowledgements}
\vspace*{1.0ex}

This work is supported by the National Natural Science Foundation of China with Grant No. 11205125.

\end{document}